\documentclass[11pt]{article}
\usepackage{amsmath,xspace,graphicx,amssymb}
\usepackage{mathptmx,helvet,courier}
\usepackage{bm}
\usepackage{setspace}
\usepackage[sort,round]{natbib}

\def\hlinefill{\leaders\hrule height3pt depth-2.5pt\hfill}
\def\emrule{\thinspace\hbox to .75em{\hlinefill}\thinspace}
\makeatletter
\def\@oddhead{\hbox{\textit{Johnson \& Goodman}} \hfil
\hbox{\textit{Jointly Poisson Processes}}}
\def\@oddfoot{\hbox{} \hfil \thepage \hfil \hbox{\today}}
\def\fnum@figure{\textbf{Figure \thefigure}}
\def\fnum@table{\textbf{Table \thetable}}

\long\def\@makecaption#1#2{%
  \vskip\abovecaptionskip
  \advance\baselineskip -2pt
  \sbox\@tempboxa{#1: #2}%
  \ifdim \wd\@tempboxa >\hsize
    {\small #1: #2}\par
  \else
    \global \@minipagefalse
    \hbox to\hsize{\hfil\box\@tempboxa\hfil}%
  \fi
  \vskip\belowcaptionskip}
\renewcommand{\section}{\@startsection {section}{1}{\z@}%
                                   {-.75ex \@plus -1ex \@minus -.2ex}%
                                   {.75ex \@plus.2ex}%
                                   {\reset@font\large\bfseries}}
\renewcommand{\subsection}{\@startsection{subsection}{2}{\z@}%
                                     {-.5ex\@plus -.4ex \@minus -.2ex}%
                                     {.5ex \@plus .2ex}%
                                     {\reset@font\normalsize\bfseries}}
\renewcommand{\subsubsection}{\@startsection{subsubsection}{3}{\z@}%
                                     {-1ex\@plus -.5ex \@minus -.2ex}%
                                     {1.5ex \@plus .2ex}%
                                     {\reset@font\normalsize\bfseries}}
\renewcommand{\paragraph}{\@startsection{paragraph}{4}{\z@}%
                                    {.5ex \@plus1ex \@minus.2ex}%
                                    {-.5em}%
                                    {\reset@font\normalsize\bfseries}}
\renewcommand{\subparagraph}{\@startsection{subparagraph}{4}{\parindent}%
                                       {1ex \@plus1ex \@minus 2ex}%
                                       {-1em}%
                                      {\reset@font\normalsize\bfseries}}

\def\@sect#1#2#3#4#5#6[#7]#8{\ifnum #2>\c@secnumdepth
     \def\@svsec{}\else
     \refstepcounter{#1}\edef\@svsec{\csname the#1\endcsname\hskip 1em }\fi
     \@tempskipa #5\relax
      \ifdim \@tempskipa>\z@
        \begingroup #6\relax
          \@hangfrom{\hskip #3\relax\@svsec}{\interlinepenalty \@M #8\par}%
        \endgroup
       \csname #1mark\endcsname{#7}\addcontentsline
         {toc}{#1}{\ifnum #2>\c@secnumdepth \else
                      \protect\numberline{\csname the#1\endcsname.}\fi
                    #7}\else
        \def\@svsechd{#6\hskip #3\@svsec #8\csname #1mark\endcsname
                      {#7}\addcontentsline
                           {toc}{#1}{\ifnum #2>\c@secnumdepth \else
                             \protect\numberline{\csname the#1\endcsname}\fi
                       #7}}\fi
     \@xsect{#5}}
\makeatother
\newcommand{\vect}[1]{\ensuremath{\mathbf{#1}}}
\newcommand{\gvect}[1]{\bm{#1}}
\newcommand{\mat}[1]{\ensuremath{\mathbf{\MakeUppercase{#1}}}}

\newcommand{\definedas}{\mathrel{\overset{\Delta}{=}}}

\newcommand{\pdf}{p}

\newcommand{\pmf}{\MakeUppercase{\pdf}}
\newcommand{\charfcn}{\Phi}
\newcommand{\mgf}{\charfcn}
\newcommand{\pgf}{G}
\newcommand{\pgfarg}{u}
\newcommand{\pgfargv}{\vect{\pgfarg}}
\newcommand{\param}{\theta}

\newcommand{\paramv}{\gvect{\param}}
\newcommand{\corr}{\rho}
\newcommand{\hocorr}[1]{\corr^{(#1)}}
\newcommand{\corrtwo}{\hocorr{2}}
\newcommand{\corrthree}{\hocorr{3}}

\newcommand{\half}{\frac{1}{2}}
\newcommand{\functionop}[1]{\textsf{#1}}
\newcommand{\E}{\functionop{E}}
\newcommand{\cov}{\functionop{cov}}
\newcommand{\var}{\functionop{var}}

\newcommand{\impulse}{\delta}

\newcommand{\construct}{\mat{A}}
\newcommand{\ratesym}{\lambda}

\newcommand{\ratesymc}{\nu}
\newcommand{\ratesymcv}{\gvect{\ratesymc}}

\newcommand{\insigsym}{x}
\newcommand{\insig}{\MakeUppercase{\insigsym}}

\newcommand{\insigsymv}{\vect{\insigsym}}
\newcommand{\insigv}{\vect{\insig}}
\newcommand{\outsigsym}{y}
\newcommand{\outsig}{\MakeUppercase{\outsigsym}}

\newcommand{\outsigv}{\vect{\outsig}}

\newcommand{\bbsym}{B}

\newcommand{\bbcountv}[1]{\vect{\bbsym}_{#1}}
\newcommand{\ppsym}{N}
\newcommand{\ppcount}[1]{{\ppsym}_{#1}}
\newcommand{\ppcountv}[1]{\vect{\ppsym}_{#1}}
\newcommand{\dinterval}{\Delta t}

\newcommand{\popindex}{m}
\newcommand{\popsize}{\MakeUppercase{\popindex}}
\newcommand{\bbindex}{l}
\newcommand{\bbsize}{\MakeUppercase{\bbindex}}

\newcommand{\nth}{^{\textrm{th}}}
\newcommand{\covm}{\gvect{\Sigma}}

\newcommand{\stddev}{\sigma}

\newcommand{\approach}{\rightarrow}

\begin{document}
\begin{center}
\textbf{\large Jointly Poisson processes}\\
\smallskip
\textit{Don H. Johnson and Ilan N. Goodman}\\
Electrical \& Computer Engineering Department, MS380\\
Rice University \\
Houston, Texas 77005--1892\\
\textit{\{dhj,igoodman\}}@rice.edu\\[6pt]
\end{center}
\sloppy

\begin{abstract}
\par\noindent
What constitutes jointly Poisson processes remains an unresolved issue. This report reviews the current state of the theory and indicates how the accepted but unproven model equals that resulting from the small time-interval limit of jointly Bernoulli processes. One intriguing consequence of these models is that jointly Poisson processes can only be positively correlated as measured by the correlation coefficient defined by cumulants of the probability generating functional.
\end{abstract}

\onehalfspacing
\section{Introduction}
To describe spike trains mathematically, particularly those that do not produce deterministic sequences of spikes, point process models are usually employed.
From a mathematical viewpoint, the Poisson process is the simplest and therefore the model that has yielded the most results.
Here, events occur randomly at a rate given by some function $\ratesym(t)$ with no statistical dependence of one event's occurrence on the number and the timing of other events.
Unfortunately, Poisson processes cannot accurately describe spike trains because of absolute and relative refractory effects.
Here, the occurrence of a spike influences when the next one occurs.
Some spike trains deviate even more from the Poisson model, with several spikes affecting subsequent ones in complicated ways.
Modeling these falls under the realm of non-Poisson processes, which in many cases makes it very difficult to obtain analytic results.
Consequently, the Poisson model is used to obtain predictions about the character of the spike train, like its information capacity, that are understood not to be precisely accurate for any realistic neural recording.
In some cases, the Poisson process can be used to obtain bounds on performance that can be used as well-established guideposts for neural behavior.

When it comes to population models, in which several neurons presumably jointly encode information, we lack even a Poisson model for all but the simplest cases:
the component point processes are either statistically independent or conditionally independent.
Data show more complicated behavior since cross-correlation functions often show  correlations among members of a population.
Consequently, what is the generalization of the single Poisson process description to what could be termed the jointly Poisson model.
Here, we seek to describe the joint statistics for several processes, each of which is Poisson (i.e., the marginal processes are Poisson).

\section{Infinite Divisibility}
From a probabilistic standpoint, specifying a unique joint probability distribution that has specified marginal distributions is ill-posed, since many joint distributions could conceivably work.
The easiest way to show the ill-posed nature of this problem is to consider the situation for Gaussian random variables.
A set of random variables $\{\insig_1,\dots,\insig_\popsize\}$ is said to be jointly Gaussian if the joint probability density has the form
\[
\pdf_{\insigv}(\insigsymv)=\frac{1}{|2\pi \covm|^{1/2}} \exp\left\{-\frac{(\insigsymv-\vect{m})'\covm^{-1}(\insigsymv-\vect{m})}{2}\right\},\;\insigv=\{\insig_1,\dots,\insig_\popsize\}
\]
Here, $\covm$ is the covariance matrix, $|\cdot|$ represents the matrix determinant, $\vect{m}$ is the vector of means and $\insigsymv'$ represents the transpose of the vector $\insigsymv$.
Each of the random variables has a Gaussian marginal probability distribution.
One can also find a joint distribution \emph{not} of this form that also has Gaussian marginals.
For example, consider the two-dimensional case ($N=2$) when the means are zero.
Let the joint distribution be as written as above, but defined to be zero in the first and third quadrants.
To obtain a valid joint distribution, we must multiply the above formula by two so that the total probability obtained by integration is one.
This joint distribution yields marginal distributions no different from the jointly Gaussian case, but the random variables are \emph{not} jointly Gaussian because the joint distribution does not have the form written above.

What makes the jointly Gaussian random vector special is the property of \emph{infinite divisibility}:
the random vector can be expressed as a sum of an arbitrary number of statistically independent random vectors~\citep{Daley-Vere-Jones-88}.
The probability distribution of the sum is the convolution of the individual probability distributions.
Consequently, infinite divisibility demands that a probability distribution be expressed as the $n$-fold convolution of a density with itself.
In special cases, like the Gaussian and the Poisson, each of the constituent random vectors has the same distributional form (i.e., they differ only in parameter values) as do their sum.

The characteristic function provides a more streamlined definition of what what infinite divisibility means.
The characteristic function of a random vector $\insigv$ is defined to be
\[
\charfcn_{\insigv}(j\pgfargv) \definedas \int \pdf_{\insigv}(\insigsymv) e^{j\pgfargv'\insigsymv}\,d\insigsymv\;.
\]
The characteristic function of a sum of statistically independent random vectors is the product of the individual characteristic functions.
\[
\charfcn_{\outsigv}(j\pgfargv)=\prod_{i=1}^n \charfcn_{\insigv_i}(j\pgfargv) \quad \outsigv=\sum_{i=1}^n \insigv_i
\]
Infinite divisibility demands that $\bigl[\charfcn_{\outsigv}(j\pgfargv)\bigr]^{1/n}$ also be a characteristic function for any positive integer value of $n$.
If we express a characteristic function parametrically as $\charfcn(j\pgfargv;\paramv)$, with $\paramv$ denotes the probability distribution's parameters, the Gaussian case is special in that $\bigl[\charfcn_{\outsigv}(j\pgfargv;\paramv)\bigr]^{1/n}=\charfcn_{\outsigv}(j\pgfargv;\paramv/n)$.
For the jointly Gaussian case, these parameters are the mean and covariance matrix.
\[
\charfcn_{\insigv_i}(j\pgfargv;\vect{m}_i,\covm_i)=\exp\left\{j\pgfargv'\vect{m}_i-\pgfargv'\covm_i\pgfargv/2\right\}
\]
Dividing these parameters by $n$ does not affect the viability of the underlying Gaussian distribution, which makes it an infinitely divisible random vector.
The example given above of a bivariate distribution having Gaussian marginals is not infinitely divisible as its characteristic function does not have this property.

In the point process case, a single Poisson process is easily seen to be infinitely divisible since the superposition of Poisson processes is also Poisson.
We must modify the just-presented mathematical formalism involving characteristic functions because we have a random process, not a random vector.
The \emph{probability-generating functional} is defined as
\[
\pgf[\pgfarg(t)] \definedas \E\left[\exp\left\{\int\log \pgfarg(t)\,d\ppcount{t}\right\}\right]\;,
\]
where the transform variable $\pgfarg(t)$ is a real-valued function of time and $\ppcount{t}$ is the point process's counting function (the number of events that have occurred prior to time $t$).
It has similar properties to the moment-generating function with one notable exception:
it has no ``inverse transform.''
However, the moment-generating function for the total number of counts in the interval implicit in the integral can be found from the probability generating function with the substitution $\pgfarg(t)\rightarrow z$.
Finding the probability distribution that underlies the expected value in the above formula requires a special series expansion.
Interesting quantities, like moments can be found from the probability-generating functional by evaluating derivatives of its logarithm.
For example, the formal derivative with respect to $\pgfarg(\cdot)$ and evaluating the result at $\pgfarg(\cdot)=1$ yields the expected value.
\begin{align*}
\frac{d\log\pgf[\pgfarg(t)]}{d\pgfarg(t)} &= \frac{1}{\pgf[\pgfarg(t)]}\left.\E\left[\int \frac{1}{\pgfarg(t)}\,d\ppcount{t}\,\exp\left\{\int\log \pgfarg(t)\,d\ppcount{t}\right\} \right]\right|_{\pgfarg(t)=1} = \E\left[\int d\ppcount{t}\right]\\
\frac{d\log\pgf[\pgfarg(t)]}{d\pgfarg(t_0)} &= \left.\frac{1}{\pgf[\pgfarg(t)]}\E\left[ \frac{1}{\pgfarg(t_0)}d\ppcount{t_0} \,\exp\left\{\int\log \pgfarg(t)\,d\ppcount{t}\right\} \right]\right|_{\pgfarg(t)=1} = \E\left[d\ppcount{t_0}\right]
\end{align*}
The first of these is the total variation with respect to $\pgfarg(t)$ and yields the expected number of events over the interval spanned by the integral.
The second is the derivative at the time instant $t_0$, which yields the expected value of the process at that time instant.

Despite not being easily able to determine the probability distribution, showing infinite divisibility can be seen by inspection just as with characteristic functions.
For a Poisson process, the probability-generating functional has the special form
\[
\pgf[\pgfarg(t)]=\exp\left\{\int\bigl(\pgfarg(t)-1\bigr)\ratesym(t)\,dt\right\}
\]
To show infinite divisibility, we note that the only ``parameter'' of a Poisson process is its instantaneous rate function $\ratesym(t)$.
As the product of probability-generating functionals for Poisson processes yields the same form with the total rate equaling the sum of the component rates, the Poisson process is infinitely divisible.

What we seek here is a description of the joint probability distribution of several marginal Poisson processes so that the vector of Poisson processes is infinitely divisible.
We exhibit here what the probability generating functional for an infinitely divisible vector of Poisson processes must be and show how to use this quantity to derive some of its properties.
In particular, we show that they can be constructed in a stereotypical way that elucidates the cross-correlation behavior required of jointly Poisson processes.
Somewhat surprisingly, the range of correlation structures is quite limited, with values for the correlation parameters tightly intertwined with each other and with the dimensionality of the vector process.
In particular, pairwise correlation coefficients cannot be negative for any pair and must decrease as the dimension increases.

\section{Jointly Poisson Processes}
The probability-generating functional for several point processes considered jointly has the simple form
\begin{equation}\label{eq:pgfdef}
\pgf^{(\popsize)}[\pgfargv(t)]\definedas \E\left[\exp\left\{\sum_{\popindex=1}^{\popsize}\int\log \pgfarg_{\popindex}(t)\,d\ppcount{\popindex,t}\right\}\right]
\end{equation}
where the expected value is computed with respect to the joint distribution of the point processes, which is the quantity we seek.
The probability-generating functional of component process $j$ can be found from this formula by setting $\pgfarg_i(t)=1$, $i\ne j$.
If the processes are statistically independent, their joint probability functional equals the product of the marginal functionals.
If the processes are added, the probability generating functional of the result equals the joint functional evaluated at a common argument:
$\pgf[\pgfarg(t)]=\pgf^{(\popsize)}[\pgfarg(t),\pgfarg(t),\dots,\pgfarg(t)]$.
These properties generalize those of moment generating functions.
Furthermore, cross-\emph{covariance} between two processes, $i$ and $j$ say, can be found by evaluating the second mixed partial of the log joint probability-generating functional:
\begin{align*}
\left.\frac{\partial^2\log\pgf^{(\popsize)}[\pgfargv(t)]}{\partial \pgfarg_i(t)\partial\pgfarg_j(t)}\right|_{\pgfargv(t)=\vect{1}} &=\E\left[\int d\ppcount{i,t}\int d\ppcount{j,t}\right]-\E\left[\int d\ppcount{i,t}\right]\cdot\E\left[\int d\ppcount{j,t}\right]\\
\left.\frac{\partial^2\log\pgf^{(\popsize)}[\pgfargv(t)]}{\partial \pgfarg_i(t_i)\partial\pgfarg_j(t_j)}\right|_{\pgfargv(t)=\vect{1}} &= 
\E\left[d\ppcount{i,t_i}d\ppcount{j,t_j}\right]-\E\left[d\ppcount{i,t_i}\right]\cdot\E\left[d\ppcount{j,t_j}\right]
\end{align*}
Again, the first expression gives the cross-covariance of counts while the second gives the cross-covariance between the processes $i$, $j$ at the times $t_i, t_j$.

Over thirty years ago, the probability-generating functional of two marginally Poisson processes that satisfied the infinite-divisibility condition was shown to have the unique form~\citep{Milne-74}
\begin{multline}\label{eq:pgftwo}
\pgf^{(2)}[\pgfarg_1(t),\pgfarg_2(t)]=\exp\left\{\int\bigl(\pgfarg_1(t)-1\bigr)\ratesymc_1(t)\,dt + \int\bigl(\pgfarg_2(t)-1\bigr)\ratesymc_2(t)\,dt\right.\\
\left.+\int\!\!\int\bigl(\pgfarg_1(s)\pgfarg_2(t)-1\bigr)\ratesymc_c(\alpha,\beta)\,d\alpha\,d\beta\right\}\;.
\end{multline}
This joint probability-generating functional is easily interpreted.
First of all, by setting $\pgfarg_2(t)=1$, we obtain the marginal probability-generating functional of process~1, showing that it is a Poisson process having an instantaneous rate of $\ratesymc_1(t)+\int\ratesymc_c(t,\beta)\,d\beta$.
Similarly, process~2 is also Poisson with a rate equal to $\ratesymc_2(t)+\int\ratesymc_c(\alpha,t)\,d\alpha$.
Also, setting $\ratesymc_c(s,t)=0$ results in the product of the marginal probability-generating functionals, corresponding to the case in which the processes are statistically independent.
Thus, the ``common rate'' $\ratesymc_c(\alpha,\beta)$ represents a joint rate variation that induces statistical dependence between the processes.
The simplest example is ${\ratesymc_c(\alpha,\beta)=\ratesymc_c(\beta)\impulse(\alpha-\beta)}$, indicating an instantaneous correlation at each moment in time.
The resulting dependence term in the probability generating functional equals
\[
\int\!\!\int\bigl(\pgfarg_1(\alpha)\pgfarg_2(\beta)-1\bigr)\ratesymc_c(\alpha,\beta)\,d\alpha\,d\beta=\int\bigl(\pgfarg_1(t)\pgfarg_2(t)-1\bigr)\ratesymc_c(t)\,dt\;.
\]

Statistically dependent Poisson processes having an infinitely divisible joint probability distribution can be simply constructed by adding to statistically independent Poisson processes having rates $\ratesymc_1(t)$ and $\ratesymc_2(t)$\emrule what we call the \emph{building-block} processes\emrule a common Poisson process having rate $\ratesymc_c(t)$ that is statistically independent of the others.
This way of constructing jointly Poisson processes amounts to the construction described by Holgate~\citep{Holgate-64}.
An allowed variant is to delay the common process when it is added to one but not the other building-block process.
Here, $\ratesymc_c(\alpha,\beta)=\ratesymc_c(\beta)\impulse\bigl(\alpha-(\beta-t_0)\bigr)$.
In this way, correlation can occur at a time lag other than zero, but still only at a single point.

More generally, $\ratesymc_c(s,t)$ depends on its arguments in different ways that do not lead to a simple superposition of building-block Poisson processes.
Using the probability generating function, you can show that the cross-covariance function between the two constructed processes equals the common rate:
$\cov\left[d\ppcount{1,t_1},d\ppcount{2,t_2}\right]=\ratesymc_c(t_1,t_2)$.
One would think that many common cross-covariances could be described this way.
However, several important constraints arise.
\begin{itemize}
\item
Cross-covariances must be non-negative.
This condition arises because the common rate must be non-negative so that a valid probability generating functional results.
\item
For the constructed processes to be jointly (wide-sense) stationary, we must have constant rates and a cross-covariance function that depends only on the time difference.
Here, the latter constraint means $\ratesymc_c(s,t)=f(|s-t|)$.
Milne and Westcott~\citep{Milne-Westcott-72} give more general conditions for the common rate function to be well-defined.
Thus, correlation can extend continuously over some time lag domain.
Consequently, the Holgate construction does not yield all possible jointly Poisson processes.
\item
It is not clear that the joint-rate characterization extends in its full generality to more than pairs of Poisson processes~\citep{Milne-Westcott-93} because the putative probability generating functional for the marginal process has not been shown to correspond to a Poisson's probability generating functional.
However, the special case of the Holgate construction technique always works.
\end{itemize}
In sequel, we only consider jointly Poisson processes that can be constructed in Holgate's fashion as a superposition of building-block Poisson processes.

Calculating means and covariances from the probability generating functional for jointly Poisson processes is very revealing.
\renewcommand{\minalignsep}{1em}
\begin{align*}
\text{Counts:} && \E\left[\int d\ppcount{i,t}\right] &= \int \bigl(\ratesymc_i(t)+\ratesymc_c(t)\bigr)\,dt\\
&& \cov\left[\int d\ppcount{1,t},\int d\ppcount{2,t}\right] &=\int\ratesymc_c(t)\,dt\\
\text{Instantaneous:} && \E\left[d\ppcount{i,t}\right] &=  \ratesymc_i(t)+\ratesymc_c(t)\\
&& \cov\left[d\ppcount{1,t_1},d\ppcount{2,t_2}\right]&=
    \begin{cases}
    0, & t_1\ne t_2\\
    \ratesymc_c(t), & t_1=t=t_2
    \end{cases}
\end{align*}
Since the variance of a Poisson process equals its mean, we find that the second-order correlation coefficient $\corrtwo(t)$ equals
\begin{align*}
\text{Counts:} && \corrtwo(t) &=\frac{\int\ratesymc_c(t)\,dt}{\sqrt{\int\bigl(\ratesymc_1(t)+\ratesymc_c(t)\bigr)\,dt\cdot\int\bigl(\ratesymc_2(t)+\ratesymc_c(t)\bigr)\,dt}}\\
\text{Instantaneous:} && \corrtwo(t) &=
   \begin{cases}
   0, & t_1\ne t_2\\
   \frac{\ratesymc_c(t)}%
   {\sqrt{(\ratesymc_1(t)+\ratesymc_c(t))(\ratesymc_2(t)+\ratesymc_c(t))}}, & t_1=t=t_2
   \end{cases}
\end{align*}
Thus, the correlation coefficient between both the counts and the instantaneous values lies in the interval $[0,1]$, with the maximal correlation occurring in the limit of large values for the common rate.
However, note that correlation has \emph{no} temporal extent and for some particular lag:
given an event occurs in one process, it is correlated with the other process at the first process's event time and uncorrelated (statistically independent) at all others.

We can write the probability-generating functional in terms of the rates of the building-block processes, $\ratesymc_i(t)$ and $\ratesymc_c(t)$, or in terms of the rates of the constructed processes $\ratesym_i(t)=\ratesymc_i(t)+\ratesymc_c(t)$ and the correlation coefficient $\corrtwo(t)$ given above.
\begin{align}
\pgf^{(2)}[\pgfarg_1(t), \pgfarg_2(t)] &=
\exp\left\{\int\bigl(\pgfarg_1(t)-1\bigr)\ratesymc_1(t)\,dt + \int\bigl(\pgfarg_2(t)-1\bigr)\ratesymc_2(t)\,dt\right. \notag\\
&\quad\left.+\int\!\!\bigl(\pgfarg_1(t)\pgfarg_2(t)-1\bigr)\ratesymc_c(t)\,dt\right\} \notag\\
\pgf^{(2)}[\pgfarg_1(t), \pgfarg_2(t)] &= \exp\left\{\int\bigl(\pgfarg_1(t)-1\bigr)\ratesym_1(t)\,dt + \int\bigl(\pgfarg_2(t)-1\bigr)\ratesym_2(t)\,dt\right.\label{eq:pgfrates}\\
&\quad\left.+\int\!\!\bigl(\pgfarg_1(t)-1\bigr)\bigl(\pgfarg_2(t)-1\bigr)\corrtwo(t)\sqrt{\ratesym_1(t)\ratesym_2(t)}\,dt\right\}\notag
\end{align}

We can extend this type of analysis to three Poisson processes constructed from six building-block processes according to the following formulas for their rates.
\begin{align*}
\ratesym_1(t) &= \ratesymc_1(t)+\ratesymc_4(t)+\ratesymc_5(t)\\
\ratesym_2(t) &= \ratesymc_2(t)+\ratesymc_4(t)+\ratesymc_6(t)\\
\ratesym_3(t) &= \ratesymc_3(t)+\ratesymc_5(t)+\ratesymc_6(t)
\end{align*}
This generates pairwise-dependent processes with no third-order dependencies.
The covariance between any pair is expressed by the building-block process rate they share in common.
Consequently,
\[
\corrtwo_{1,2}=\frac{\ratesymc_4(t)}{\sqrt{\ratesym_1(t)\ratesym_2(t)}}\;.
\]

By letting $\ratesymc_1=\ratesymc_2\equiv\ratesymc^{(1)}$ and $\ratesymc_4=\ratesymc_5=\ratesymc_6\equiv\ratesymc^{(2)}$, we create what we term the \emph{symmetric} case, in which we have only two separately adjustable rates that arise from the six statistically independent building-block processes.
In this case, this cross-correlation simplifies to
\begin{equation}\label{eq:corrtwo}
\corrtwo=\frac{\ratesymc^{(2)}(t)}{\ratesymc^{(1)}(t)+2\ratesymc^{(2)}(t)}\le \half,\; i \ne j
\end{equation}

When a Poisson process having instantaneous rate $\ratesymc^{(3)}(t)$ is added to all three building-block processes to create third-order dependence, the correlation coefficient becomes in the symmetric case
\[
\corrtwo=\frac{\ratesymc^{(2)}(t)+\ratesymc^{(3)}(t)}{\ratesymc^{(1)}(t)+2\ratesymc^{(2)}(t)+\ratesymc^{(3)}(t)},\; i\ne j
\]
Now, as the common process's rate grows, the pairwise correlation coefficient can approach one.
If we define a third-order correlation coefficient according to
\begin{equation}\label{eq:thirdorder}
\corrthree[d\ppcount{1,t},d\ppcount{2,t},d\ppcount{3,t}] \definedas \frac{\left.\frac{\displaystyle\partial^3 \log\pgf[\pgfarg_1(t),\pgfarg_2(t),\pgfarg_3(t)]}{\displaystyle\partial \pgfarg_1(t) \partial \pgfarg_2(t) \partial \pgfarg_3(t)}\right|_{\pgfargv=\vect{1}}}{\sqrt[3]{\var[d\ppcount{1,t}]\var[d\ppcount{2,t}]\var[d\ppcount{3,t}]}}\;.
\end{equation}
For the symmetric Poisson example, the third-order correlation coefficient is easily found to be
\[
\corrthree(t)=\frac{\ratesymc^{(3)}(t)}{\ratesymc^{(1)}(t)+2\ratesymc^{(2)}(t)+\ratesymc^{(3)}(t)}
\]
Combining with the expression for the second-order correlation coefficient, we find the following bounds for the symmetric case relating the correlation quantities.
\[
0 \le \corrthree \le 2\corrtwo-\corrthree \le 1
\]
Note that this inequality chain indicates that $0\le\corrthree\le\corrtwo\le1$.
The second-order correlation can be bigger than $\half$, but only if $\corrthree$ increases as well in a manner defined by the inequality chain.

We need to extend this analysis to an arbitrary number of building block and constructed processes.
We can form an arbitrary number of infinitely divisible, jointly defined Poisson processes by extending the two- and three-process Holgate construction technique.
Given $\bbsize$ statistically independent Poisson processes, we create a population of $M$ statistically dependent Poisson processes according by superimposing $\bbsize$ building-block processes according to the \emph{construction matrix} $\construct$: $\ppcountv{t}=\construct\bbcountv{t}$.
Here, $\ppcountv{t}$  and $\bbcountv{t}$ represent column vectors of constructed and building-block Poisson processes of dimension $\popsize$ and $\bbsize>\popsize$ respectively.
The entries of the construction matrix are either $0$ or $1$.
For example, the construction matrix underlying the two- and three-process examples are
\begin{align*}
\popsize=2\colon& \construct=\begin{bmatrix}1 & 0 & 1\\ 0 & 1 & 1\end{bmatrix}\\
\popsize=3\colon& \construct=
\begin{bmatrix}
  1 & 0 & 0 & 1 & 1 & 0 & 1\\
  0 & 1 & 0 & 1 & 0 & 1 & 1\\
  0 & 0 & 1 & 0 & 1 & 1 & 1
\end{bmatrix}
\end{align*}
To introduce dependencies of all orders, $\bbsize\ge 2^{\popsize}-1$, and we concentrate on the case $\bbsize=2^{\popsize}-1$ in sequel.

The probability generating functional $\pgf^{(\popsize)}[\pgfargv(t)]$ of $\ppcountv{t}$ expressed in~\eqref{eq:pgfdef} can be written in matrix form as
\[
\pgf^{(\popsize)}[\pgfargv(t)]=\E\left[\exp\left\{\int \log \pgfargv'(t)\,d\ppcountv{t}\right\}\right]
\]
where the logarithm of a vector is defined in the \textsc{Matlab} sense (an element-by-element operation).
Because $\ppcountv{t}=\construct\bbcountv{t}$, we have
\begin{align*}
\pgf^{(\popsize)}[\pgfargv(t)]&=\E\left[\exp\left\{\int \log \pgfargv'(t)\construct\,d\bbcountv{t}\right\}\right]\\
&=\E\left[\exp\left\{\int \left(\construct'\log \pgfargv(t)\right)'\,d\bbcountv{t}\right\}\right]
\end{align*}
Each component of the vector $\construct'\log \pgfargv(t)$ expresses which combination of components of $\pgfargv(t)$ are associated with each building block process.
This combination corresponds to the constructed processes to which each building block process contributes.
Since the building block processes are statistically independent and Poisson, we have
\[
\pgf^{(\popsize)}[\pgfargv(t)] = \int \left[\exp\left\{\construct'\log\pgfargv(t)\right\}-1\right]'\ratesymcv(t)\,dt
\]
Expanding the vector notation for a moment, this result can also be written as
\begin{equation}\label{eq:pgfbb}
\pgf^{(\popsize)}[\pgfargv(t)] = \exp\left\{\sum_{\bbindex=1}^{\bbsize} \int\left(\left[\prod_{\popindex=1}^{\popsize} \pgfarg_{\popindex}^{A_{\popindex,\bbindex}}(t)\right]-1\right)\ratesymc_{\bbindex}(t)\, dt\right\}
\end{equation}
Here, $\pgfarg_{\popindex}^{A_{\popindex,\bbindex}}(t)$ means $\pgfarg_{\popindex}(t)$ raised to the $A_{\popindex,\bbindex}$ power.
In other words, if $A_{\popindex,\bbindex}=1$, the term is included;
if $A_{\popindex,\bbindex}=0$ it is not.
Thus, the probability generating functional consists of a sum of terms, one for each building block process, wherein the coefficient of each rate $\ratesymc_{\bbindex}(t)$ is the product of arguments corresponding to those constructed process building block process $\bbindex$ helped to build minus one.
This form is what equation~\eqref{eq:pgftwo} describes.

However, we need to convert this result into the form of~\eqref{eq:pgfrates} so that the role of the cumulant correlation coefficients can come to light.
We can view the cumulant moments, the mixed first partials of the logarithm of the probability generating functional, as coefficients of the multivariate Taylor series for $\log\pgf^{(\popsize)}[\pgfargv(t)]$ centered at the point $\pgfargv(t)=\vect{1}$.
Because the $\popindex\nth$ term in~\eqref{eq:pgfbb} contains only multilinear combinations of $\pgfarg_{\popindex}$, second-order and higher derivatives of these terms are zero.
Consequently, the Taylor series for $\log\pgf^{(\popsize)}[\pgfargv(t)]$ consists \emph{only} of multilinear terms having $(\pgfarg_{\popindex}-1)$ as its constituents with the cumulant moments as the series coefficients.
Consequently, the jointly Poisson process can always be written in a form generalizing~\eqref{eq:pgfrates}.
This coefficient equals
\begin{equation}\label{eq:taylorcoeff}
\left.\frac{\partial^k\log\pgf^{(\popsize)}[\pgfargv(t)]}{\partial \pgfarg_{\popindex_1}(t)\dots\partial\pgfarg_{\popindex_k}(t)}\right|_{\pgfargv(t)=\vect{1}}
=
\sum_{\bbindex=1}^{\bbsize} \left(\prod_{\popindex=\popindex_1,\ldots,\popindex_k} A_{\popindex,\bbindex}\right)\ratesymc_{\bbindex}(t)
\end{equation}
Because matrix $\construct$ has only binary-valued entries, the product $\prod_{\popindex} A_{\popindex,\bbindex}$ equals either one or zero, bringing in the $\bbindex\nth$ building block process only if it contributes to all of the constructed processes indexed by $\popindex_1,\ldots,\popindex_k$.
Note that the first partial derivative expresses the rate of each constructed process:
$\ratesym_{\popindex}(t)=\sum_{\bbindex}A_{\popindex,\bbindex}\ratesymc_{\bbindex}(t)$.

We can normalize the Taylor series coefficient to obtain cumulant correlation coefficients by dividing by the geometric mean of the constructed process rates that enter into the partial derivative shown in~\eqref{eq:taylorcoeff}.
\begin{align*}
\hocorr{k}_{\popindex_1,\ldots,\popindex_k}(t) &\definedas \frac{\left.\frac{\partial^k\log\pgf^{(\popsize)}[\pgfargv(t)]}{\partial \pgfarg_{\popindex_1}(t)\dots\partial\pgfarg_{\popindex_k}(t)}\right|_{\pgfargv(t)=\vect{1}}}{\left[\ratesym_{m_1}(t)\cdots\ratesym_{m_k}(t)\right]^{1/k}}\\
&= \frac{\sum_{\bbindex=1}^{\bbsize} \left(\prod_{\popindex=\popindex_1,\ldots,\popindex_k} A_{\popindex,\bbindex}\right)\ratesymc_{\bbindex}(t)}{\left[\sum_{\bbindex}A_{\popindex_1,\bbindex}\ratesymc_{\bbindex}(t)\cdots\sum_{\bbindex}A_{\popindex_k,\bbindex}\ratesymc_{\bbindex}(t)\right]^{1/k}}
\end{align*}
Because the numerator expresses which building block processes are in common with all the specified constructed processes, they and others are contained in each term in the denominator.
This property means that each cumulant correlation coefficient is less than one and, since rates cannot be negative, greater than or equal to zero.
Similar manipulations show that $\hocorr{k}_{\popindex_1,\ldots,\popindex_k}(t)\ge \hocorr{k+1}_{\popindex_1,\ldots,\popindex_k,\popindex_{k+1}}(t)$:
the size of the cumulant correlation coefficients cannot increase with order.

In the symmetric case, the expression for the cumulant correlation coefficients simplifies greatly.
\begin{equation}\label{eq:rhokdef}
\hocorr{k}(t)=\frac{\sum_{\bbindex=k}^\popsize \binom{\popsize-k}{\bbindex-k}\ratesymc^{(\bbindex)}(t)}{\sum_{\bbindex=1}^\popsize \binom{\popsize-1}{\bbindex-1}\ratesymc^{(\bbindex)}(t)}
\end{equation}
The denominator is the rate $\ratesym(t)$ of each constructed process and the numerator is the sum of the rates of the processes that induce the dependence of the specified order.
This result makes it easier to see that the cumulant correlation coefficients cannot increase in value with increasing order:
$0\le\corr^{(k)}(t)\le\corr^{(k-1)}(t)\le1$, $k=3,\dots,\popsize$.
Furthermore, more stringent requirements can be derived by exploiting the structure equation~\eqref{eq:rhokdef}, showing that the cumulant correlation coefficients must obey the following two relationships in the symmetric case.
\begin{equation}\label{eq:rhochain}
\begin{aligned}
&\sum_{k=2}^{\popsize}{\hocorr{k}(-1)^k\binom{\popsize-1}{k-1}} \le 1\\
&\sum_{k=\popindex}^{\popsize}{\hocorr{k}(-1)^{k+\popindex}\binom{\popsize-\popindex}{k-\popindex}} \ge 0,\quad \popindex = 2,\ldots,\popsize
\end{aligned}
\end{equation}
For example, for four jointly Poisson processes, the cumulant correlation coefficients must satisfy the inequalities
\[
\begin{gathered}
 3\corrtwo-3\corrthree+\hocorr{4}\le 1\\
\corrtwo-2\corrthree+\hocorr{4}\ge 0
\end{gathered}
\]

\section{Relations to Jointly Bernoulli Processes}
Interestingly, this form of the jointly Poisson process can be derived as the limit of the jointly Bernoulli process when the event probability becomes arbitrarily small.
First of all, a single Poisson process is defined this way, with the event probability equal to $\ratesym(t)\dinterval$.
To extend this approach to two jointly Poisson processes, we use the Sarmanov-Lancaster model for two jointly Bernoulli processes~\citep{Goodman-04}.
Letting $\insig_1, \insig_2$ be Bernoulli random variables with event probabilities $\pdf_1,\pdf_2$ respectively, the joint probability distribution is given by
\[
\pmf(\insig_1, \insig_2) =
\pmf(\insig_1)\pmf(\insig_2)\left[1+\corr\frac{(\insig_1-\pdf_1)(\insig_2-\pdf_2)}{\stddev_1\stddev_2}\right]
\]
where the standard deviation $\stddev_i$ of each random variable equals $\sqrt{\pdf_i(1-\pdf_i)}$.
The key to the derivation is to use the moment generating function, defined to be the two-dimensional $z$-transform of this joint distribution.
\[
\mgf(z_1,z_2)=\sum_{x_1}\sum_{x_2}\pmf(x_1,x_2)z_1^{x_1}z_2^{x_2}
\]
Simple calculations show that for the jointly Bernoulli distribution given above, its moment generating function is
\begin{align*}
\mgf(z_1,z_2) &= \left[(1-\pdf_1)(1-\pdf_2)+\corr\stddev_1\stddev_2\right]
+\left[\pdf_1(1-\pdf_2)-\corr\stddev_1\stddev_2\right]z_1
+\left[\pdf_2(1-\pdf_1)-\corr\stddev_1\stddev_2\right]z_2\\
&\quad+\left[\pdf_1\pdf_2+\corr\stddev_1\stddev_2\right]z_1z_2\\
&=\bigl(1+\pdf_1(z_1-1) \bigr)\bigl(1+\pdf_2(z_2-1)\bigr)+(z_1-1)(z_2-1)\corr\stddev_1\stddev_2
\end{align*}
Letting event probabilities be proportional to the binwidth $\dinterval$, we evaluate this expression to first order in the event probabilities.
Especially note that $\stddev_1\stddev_2\approx \sqrt{\ratesym_1\ratesym_2}\dinterval$ as $\dinterval\approach0$ to first order.
Therefore, we have
\begin{align*}
\mgf(z_1,z_2) \mathrel{\overset{\dinterval\approach0}{\longrightarrow}} & [1+(z_1-1)\ratesym_1\dinterval] [1+(z_2-1)\ratesym_2\dinterval]+(z_1-1)(z_2-1)\corr\sqrt{\ratesym_1\ratesym_2}\dinterval\\
=& 1+\ratesym_1\dinterval(z_1-1)+\ratesym_2\dinterval(z_2-1)+\corr\sqrt{\ratesym_1\ratesym_2}\dinterval(z_1-1)(z_2-1)
\end{align*}
Evaluating the natural logarithm and using the approximation $\log(1+x)\approx x$ for small $x$, we find that
\[
\log \mgf(z_1,z_2)\approx (z_1-1)\ratesym_1\dinterval+(z_2-1)\ratesym_2\dinterval+(z_1-1)(z_2-1)\corr\sqrt{\ratesym_1\ratesym_2}\dinterval
\]
If we sum the Bernoulli random variables in each process over a fixed time interval, say $[0,T]$, we obtain the number of events that occur in each process.
The moment generating function of this sum is the product of the individual joint moment generating functions, which means its logarithm equals the sum of the logarithms of the individual functions.
Since the number of random variables increases as the binwidth decreases (equal to $T/\dinterval$) and noting these terms are proportional to $\dinterval$, the sum becomes an integral to yield
\[
\log \mgf(N_1,N_2)= (z_1-1)\int_{0}^{T}\!\!\!\ratesym_1(t)\,dt+(z_2-1)\int_{0}^{T}\!\!\!\ratesym_2(t)\,dt+(z_1-1)(z_2-1)\int_{0}^{T}\!\!\!\corr(t)\sqrt{\ratesym_1(t)\ratesym_2(t)}\,dt
\]
If we let $\ratesym_i(t)=\ratesymc_i(t)+\ratesymc_c(t)$ and substitute~\eqref{eq:corrtwo} for the definition of the correlation coefficient, we obtain the logarithm of the probability generating functional for two jointly Poisson processes constructed using Holgate's method in which $\pgfarg_i(t)\rightarrow z_i$ as in equation~\eqref{eq:pgfrates}.

Generalizing this result is tedious but straightforward:
jointly Bernoulli processes converge in the limit of small event probabilities to jointly Poisson processes interdependent on each other at the same moment.
An interesting sidelight is the normalization of the higher order dependency terms in the Sarmanov-Lancaster expansion demanded to make the correlation coefficient in the two models agree.
In the Sarmanov-Lancaster expansion, the $k\nth$ order term has the form exemplified by
\[
\corr^{(k)}\frac{(\insig_1-\pdf_1)\cdots(\insig_k-\pdf_k)}{C_k}
\]
where $C_k$ is the normalization constant that depends on correlation order and the specific choice of random variables in the term.
Normally, Sarmanov-Lancaster expansions consist of products of orthonormal functions, which in this case would be $\prod (\insig_i-\pdf_i)/\stddev_i$.
This makes the putative normalization constant equal to $C_k=\prod \stddev_i$.
However, the higher order correlation coefficients consequent of this definition have no guaranteed domains as does $\corr^{(2)}$.
As described above, the jointly Poisson correlation coefficients defined via cumulants do have an orderliness.
Associating the two demands that correlation coefficient be defined as
\[
\corr^{(k)}\definedas \frac{\E\bigl[(\insig_1-\pdf_1)\cdots(\insig_k-\pdf_k)\bigr]}{\left(\prod_{i=1}^{k} \stddev_i^2\right)^{1/k}}
\]
The normalization $(\prod_{i=1}^k \stddev_i^2)^{1/k}$ corresponds to the geometric mean of the variances found in the definition~\eqref{eq:thirdorder} of correlation coefficients for Poisson processes.
In the context of the Sarmanov-Lancaster expansion, we have
\[
\corr^{(k)}= \corr^{(k)}\frac{\stddev_1^2\cdots\stddev_k^2}{C_k \cdot\left(\prod \stddev_i^2\right)^{1/k}}\;.
\]
Solving for $C_k$, we find that
\[
C_k = \left(\stddev_1^2\cdots\stddev_k^2\right)^{\frac{k-1}{k}}\;.
\]
Using this normalization in the Sarmanov-Lancaster expansion now creates a direct relationship between its parameters and those of the jointly Poisson probability distribution.
The inequality sets shown in~\eqref{eq:rhochain} also guarantee existence of the Sarmanov-Lancaster model~\citep{Bahadur-61}.
This change does not affect the orthogonality so crucial in defining the Sarmanov-Lancaster expansion, only the normality.

Because of the correspondence between jointly Bernoulli processes and jointly Poisson processes, we can use the limit of the Sarmanov-Lancaster expansion to represent the joint distribution of jointly Poisson processes.
In particular, we can evaluate information-theoretic quantities related to Poisson processes using this correspondence.
Since entropy and mutual information are smooth quantities (infinitely differentiable), the small-probability limit can be evaluated \emph{after} they are computed for Bernoulli processes.

\singlespacing

\end{document}